\documentclass[aps,pra,twocolumn,floats,amsfonts,superscriptaddress]{revtex4}
\usepackage{times}

\begin{document}

\title{Comment on ``Impossibility of distant indirect measurement of
  the quantum Zeno effect''}

\author{S.~Wallentowitz} 

\affiliation{Facultad de F{\'i}sica, Pontificia Universidad Cat{\'o}lica de
  Chile, Casilla 306, Santiago 22, Chile}

\author{P.E.~Toschek}

\affiliation{Institut f\"ur Laser-Physik, Universit\"at
  Hamburg, Luruper Chaussee 149, D-22761 Hamburg, Germany}

\date{October 27, 2004} 

\begin{abstract}
  In the paper by M. Hotta and M. Morikawa [Phys. Rev. A {\bf 69},
  052114 (2004)] the non-existence of the quantum Zeno effect caused
  by indirect measurements has been claimed. It is shown here that the
  pertinent proof is incorrect, and the claim unfounded.
\end{abstract}


\maketitle

The authors of the quoted paper~\cite{hotta} claim non-existence of
the impediment of the evolution of a quantum system by its indirect
measurement, known as the quantum Zeno effect (QZE). They attempt to
prove that with distant indirect measurements the final survival
probability of the object is not affected by repeated measurements. We
show that this proof is faulty, and consequently the above claim is
unfounded.

The proof involves the subdivision of the system's Hilbert space
${\cal H}_{\cal Z} \!=\! {\cal H}_{\cal C} \oplus {\cal H}_{\cal W}$
in a core-zone supspace ${\cal H}_{\cal C}$ and a wave-zone subspace
${\cal H}_{\cal W} $. By definition, throughout the time evolution a
wave-zone state remains in ${\cal H}_{\cal W}$, see Eq.~(7) of
Ref.~\cite{hotta}. This is formally expressed by the vanishing of the
operator product $\hat{P}_{\cal C} \hat{U}_{+}(t) \hat{P}_{\cal W}$,
where $\hat{U}_+(t)$ represents the unitary time-evolution operator
and
\begin{equation}
  \hat{P}_{\cal C} =  \sum_{C} |C \rangle \langle C | , \quad 
  \hat{P}_{\cal W} =  \sum_{W} |W \rangle \langle W |
\end{equation}
are the projectors into core and wave zone, respectively.  However, a
core-zone state $|C\rangle \!\in\! {\cal H}_{\cal C}$ is said to
decay, with finite probability, into a wave-zone state [see
Eqs~(19)--(21) of Ref.~\cite{hotta}]. Such an evolution requires the
operator product $\hat{P}_{\cal W} \hat{U}_{+}(t) \hat{P}_{\cal C}$ to
be \textit{non-vanishing}. 

We prove here that, in contrast to this assumption, this operator
product exactly vanishes:
\begin{equation}
  \label{eq:prove}
 \hat{P}_{\cal W} \hat{U}_{+}(t) \hat{P}_{\cal C} \equiv 0 .
\end{equation}
For the proof we start with the explicit expression for the unitary
time-evolution operator
\begin{equation}
  \label{eq:U}
  \hat{U}_{+}(t) =  {\cal T} \exp \!\left[ -\frac{i}{\hbar} \int_0^t \! dt'
  \, \hat{H}(t') \right]
\end{equation}
where ${\cal T}$ indicates the proper time ordering.
Equation~(\ref{eq:U}) may be expanded in a series of powers of the
Hamiltonian $\hat{H}(t)$ as
\begin{equation}
  \label{eq:series}
  \hat{U}_+(t) = \sum_{n=0}^\infty \left( - \frac{i}{\hbar} \right)^n 
  \int_0^t \! dt_n \ldots \int_0^{t_2} \! dt_1 \, \hat{H}(t_n) \ldots
  \hat{H}(t_1) . 
\end{equation}
According to Eq.~(7) of Ref.~\cite{hotta} the relation $\hat{U}_+(t) |
W \rangle \!\in\! {\cal H}_{\cal W}$ holds, which upon insertion of
Eq.~(\ref{eq:series}) can be shown to imply $\hat{H}(t) | W \rangle
\!\in\! {\cal H}_{\cal W}$.  From this relation it follows that
consequently $\langle W | \hat{H}^\dagger(t) \!=\! \langle W |
\hat{H}(t) \!\in\! {\cal H}_{\cal W}^\ast$, where ${\cal H}_{\cal
  W}^\ast$ denotes the dual Hilbert space of ${\cal H}_{\cal W}$, and
thus one concludes that the following relation holds
\begin{equation}
  \label{eq:W-left}
  \langle W | \hat{U}_{+}(t) = \sum_{W'} \langle W | \hat{U}_+(t) | W'
  \rangle \langle W' | \in {\cal H}_{\cal W}^\ast . 
\end{equation}
This relation, being absent in Ref.~\cite{hotta}, is the central
ingredient that allows us to prove now Eq.~(\ref{eq:prove}).

Equation~(\ref{eq:W-left}) [together with Eq.~(7) of
Ref.~\cite{hotta}] can now be used to prove that the commutator
$[\hat{P}_{\cal W}, \hat{U}_+(t) ]$ vanishes:
\begin{eqnarray}
  \label{eq:comm}
  [ \hat{P}_{\cal W}, \hat{U}_{+}(t) ] & = & 
  \sum_W \left[ |W\rangle \langle W| \, \hat{U}_{+}(t) 
    - \hat{U}_{+}(t) \, |W \rangle \langle W | \right] \nonumber \\
  & = & \sum_{W,W'} \Big[ |W \rangle \langle W |
  \hat{U}_{+}(t) |W'\rangle \langle W' | \nonumber \\
  & & 
  \qquad - \, |W'\rangle\langle W' | \hat{U}_{+}(t) |W\rangle \langle
  W| \Big] \nonumber \\
  & \equiv & 0 .
\end{eqnarray}
Of course then also
\begin{equation}
  \label{eq:trivial}
  [ \hat{P}_{\cal W}, \hat{U}_+(t) ] \hat{P}_{\cal W} \equiv 0 ,
\end {equation}
which allows us to perform one last step by formally 
inserting $\hat{P}_{\cal W} \!=\! \hat{1} \!-\! \hat{P}_{\cal C}$  to  yield 
\begin{equation}
  [ \hat{P}_{\cal W}, \hat{U}_{+}(t) ] - \hat{P}_{\cal W}
  \hat{U}_{+}(t) \hat{P}_{\cal C} + \hat{U}_{+}(t) \hat{P}_{\cal W}
  \hat{P}_{\cal C} \equiv 0 .
\end {equation}
Since both the first term, see Eq.~(\ref{eq:comm}), and the third term 
(cf. $\hat{P}_{\cal W} \hat{P}_{\cal C} \!\equiv\!0$) are zero, also
the second term necessarily vanishes:
\begin{equation}
  \hat{P}_{\cal W} \hat{U}_{+}(t) \hat{P}_{\cal C} \equiv 0 ,
\end {equation}
and thus Eq.~(\ref{eq:prove}) has been proved.

This equality proves \textit{no transition} to occur
\textit{from core zone to wave zone} either. The subspaces are
completely  separated, and the probability of finding that the
observed state belongs to $\cal{H}_{\cal W}$
[see Eq.~(20) of Ref.~\cite{hotta}] vanishes: $p_{1} \!=\! 0$. 
In terms of measurements, the system and meter are disconnected, 
and the QZE is excluded on trivial grounds.

The correct analysis requires consideration of the formal structure of
$\hat{U}_+(t)$ in terms of the Hamiltonian $\hat{H}(t)$, whose result
is the above Eq.~(\ref{eq:W-left}). Rather than the time-evolution
operator's unitarity, $U^\dagger_{+}(t) \!=\! U_{-}(t)$, its left-hand 
operation warrants time reversal and the correct result, for
$\hat{H}(t)$ being an arbitrary time-dependent Hamiltonian, including
the special cases of a constant or an even function of time.

\end{document}